\documentclass[10pt,reprint,aip,amsmath,amssymb]{revtex4-1}
\usepackage{graphicx}
\input epsf

\newcommand{\Order}{\mbox{$\mathcal{O}$}} 
\newcommand\Alfven{Alfv\'en }
\newcommand\Alfvenic{Alfv\'enic }
\newcommand{\V}[1]{\mathbf{#1}} 
\newcommand{\T}[1]{\texttt{#1}} 
 
\newcommand{\zhat}{\mbox{$\hat{\mathbf{z}}$}} 
\newcommand{\xhat}{\mbox{$\hat{\mathbf{x}}$}} 
\newcommand{\yhat}{\mbox{$\hat{\mathbf{y}}$}} 
\newcommand{\figref}[1]{Figure~\ref{#1}}
\newcommand{\secref}[1]{\S\ref{#1}}

\begin{document}


\title{\Alfven Wave Collisions, The Fundamental 
Building Block of Plasma Turbulence II: Numerical Solution} 



\author{K.~D. Nielson}
\author{G.~G. Howes}
\email[]{gregory-howes@uiowa.edu}
\affiliation{Department of Physics and Astronomy, University of Iowa, Iowa City, 
Iowa 52242, USA.}

\author{W.~Dorland}
\affiliation{Department of Physics, University of Maryland, College Park, 
Maryland 20742-3511, USA.}


\date{\today}

\begin{abstract}
This paper presents the numerical verification of an asymptotic
analytical solution for the nonlinear interaction between
counterpropagating \Alfven waves, the fundamental building block of
astrophysical plasma turbulence. The analytical solution, derived in
the weak turbulence limit using the equations of incompressible MHD,
is compared to a nonlinear gyrokinetic simulation of an \Alfven wave
collision. The agreement between these methods signifies that the
incompressible solution satisfactorily describes the essential
dynamics of the nonlinear energy transfer, even under the weakly
collisional plasma conditions relevant to many astrophysical
environments.
\end{abstract}


\maketitle 


\section{Introduction}

Plasma turbulence impacts the evolution of many space and
astrophysical environments of interest, primarily by mediating the
transfer of energy from large-scale motions to sufficiently small
scales that the turbulence can be dissipated, ultimately leading to
heating of the plasma. The turbulent cascade of energy in
astrophysical plasmas is driven by nonlinear interactions between
counterpropagating \Alfven waves, commonly called \Alfven wave
collisions.  In our companion paper, Howes and
Nielson,\cite{Howes:2013a} hereafter Paper I, we discuss the
properties of this fundamental building block of astrophysical plasma
turbulence and present an asymptotic analytical solution for the
nonlinear evolution of the interaction between two counterpropagating
\Alfven waves using the incompressible magnetohydrodynamics (MHD)
equations in the weakly nonlinear limit.  The primary aim of this
paper is to present a numerical verification of this analytical
solution.

The incompressible MHD solution for the nonlinear interaction between
two counterpropagating \Alfven waves derived in Paper I provides
useful insight into the nature of the turbulent cascade of energy in
magnetized astrophysical plasmas. Many such astrophysical plasmas,
however, do not satisfy the conditions necessary for the validity of
the incompressible MHD equations, calling into question the
applicability of such an idealized solution. The standard MHD
approximation is valid for non-relativistic ($v_{t i} /c \ll1 $),
large-scale ($k \rho_i \ll 1$), and low-frequency ($\omega \ll
\Omega_i$) motions under strongly collisional plasma conditions
($\omega \ll \nu$). In addition, incompressibility requires a sound
speed much larger than the \Alfven speed,\cite{Howes:2013a} or the
large plasma beta limit ($\beta \gg 1$).  Although the inertial range
turbulent dynamics of many space and astrophysical plasmas indeed
satisfy the first three conditions, such plasmas are often weakly
collisional, $\omega \gg \nu$, and may have a low or order-unity
plasma beta, $\beta \lesssim 1$. We hypothesize here that, \emph{even
under weakly collisional conditions with order-unity plasma beta, the
nonlinear dynamics between counterpropagating \Alfven waves that
underlies the turbulent cascade of energy remains well described by
the incompressible MHD solution derived in Paper I}. The secondary aim
of this paper is to test this hypothesis by directly comparing the
incompressible MHD analytical solution from Paper I to the nonlinear
numerical evolution described by
gyrokinetics,\cite{Frieman:1982,Howes:2006,Schekochihin:2009} a
formalism that rigorously describes the kinetic plasma dynamics in the
limit of weak collisionality and order-unity plasma beta. This
numerical verification establishes the applicability of the
qualitative picture of \Alfven wave collisions described in Paper I to
turbulence in realistic space and astrophysical plasma environments of
interest.

The gyrokinetic numerical method used to simulate the collision
between the \Alfven waves is described in \secref{sec:method}. In
\secref{sec:numsol}, the numerical solution is presented, with a 
detailed verification of the predicted time evolution for both the
complex Elsasser potentials and the real electromagnetic fields. The
results are discussed and conclusions drawn in \secref{sec:discuss}.

\section{Numerical Method}
\label{sec:method}
To verify the analytical solution derived in Paper I, we perform a
nonlinear simulation of the collision between two counterpropagating
\Alfven waves using \T{AstroGK}, the Astrophysical Gyrokinetics 
Code, developed specifically to study kinetic turbulence in
astrophysical plasmas. Although it may seem an unusual choice, the use
of a gyrokinetic code to verify an analytical solution derived in the
framework of incompressible MHD is a deliberate one.

First, recall from Paper I that the analytical solution presented
there is formally valid only in the anisotropic limit, $k_\perp \gg
k_\parallel$, a limit that arises naturally in magnetized plasma
turbulence, hereafter denoted the \emph{anisotropic limit}. In this
anisotropic limit, the low-frequency kinetic dynamics of a weakly
collisional plasma is rigorously described by the gyrokinetic system
of equations.\cite{Frieman:1982,Howes:2006,Schekochihin:2009} In the
limit of perpendicular scales much larger than the ion Larmor radius,
$k_\perp
\rho_i \ll 1$, it has been shown that the \Alfvenic dynamics in a 
weakly collisional plasma is formally governed by the much more simple
equations of reduced MHD.\cite{Schekochihin:2009} As shown in Paper I,
in the anisotropic limit $k_\perp\gg k_\parallel$, the incompressible
MHD description of the \Alfvenic dynamics is equivalent to the reduced
MHD description.  Therefore, we expect that the analytical solution
derived in Paper I is valid description of the \Alfvenic dynamics
described by gyrokinetics in the limit $k_\perp \rho_i \ll 1$,
hereafter denoted the \emph{MHD limit}.

Second, we choose to verify the analytical solutions with a
gyrokinetic code to test our hypothesis that the essential dynamical
behavior of the turbulent energy cascade in astrophysical plasma
environments is well described using the simplified framework of
incompressible MHD. Many of the fundamental concepts that form the
foundation of modern theories for plasma turbulence have been derived
in the context of incompressible MHD, so it is important to determine
if these properties persist under more general plasma conditions in
order to establish the applicability of these idealized theoretical
concepts to turbulence in real space and astrophysical plasmas.  As
discussed in the introduction, such astrophysical plasmas are
frequently found to be weakly collisional, so a kinetic description of
the turbulent dynamics is formally required.  In the anisotropic limit
that arises naturally in magnetized plasma turbulence, the
low-frequency kinetic dynamics of the turbulence is properly captured
by a gyrokinetic
description.\cite{Howes:2006,Howes:2008b,Schekochihin:2009} Therefore,
by using a gyrokinetic code to validate our incompressible MHD
solutions for the dynamics of the nonlinear interaction between
counterpropagating \Alfven waves, we can determine if the kinetic
dynamics of the turbulent energy cascade in weakly collisional space and
astrophysical plasmas is adequately described by the much more simple
and analytically tractable equations of incompressible MHD. Moreover,
in the future, as we extend our investigation to the smaller scales
$k_\perp \rho_i\gtrsim 1$ where the linear wave physics becomes
dispersive, we can explore how the turbulent energy transfer changes
character as it transitions from a cascade of nondispersive \Alfven
waves to a cascade of dispersive kinetic \Alfven
waves.\cite{Howes:2006,Howes:2008a,Howes:2008b,Schekochihin:2009,Howes:2011a,Howes:2011b}

\subsection{Numerical Code Description}
We use \T{AstroGK}, the Astrophysical Gyrokinetics Code, to simulate
the nonlinear evolution of the collision between counterpropagating
\Alfven waves. A detailed description of the algorithms in the code and the results of linear and nonlinear benchmarks are presented in Numata \emph{et
al.}\cite{Numata:2010}, so we give here only a brief overview.

\T{AstroGK} evolves the perturbed gyroaveraged distribution
function $h_s(x,y,z,\lambda,\varepsilon)$ for each species $s$, the
scalar potential $\varphi$, parallel vector potential $A_\parallel$,
and the parallel magnetic field perturbation $\delta B_\parallel$
according to the gyrokinetic equation and the gyroaveraged Maxwell's
equations.\cite{Frieman:1982,Howes:2006} The gyroaveraging procedure
reduces the three-dimensional velocity space to the components
parallel and perpendicular to the equilibrium magnetic field,
$v_\parallel$ and $v_\perp$; in the code, a complementary
representation of velocity space is chosen using the pitch angle
$\lambda=v_\perp^2/v^2$ and the energy $\varepsilon=v^2/2$. The domain
is a periodic box of size $L_{\perp }^2 \times L_{\parallel }$,
elongated along the straight, uniform mean magnetic field, $\V{B}_0=B_0 \zhat$. Note
that, in the gyrokinetic formalism, all quantities may be rescaled to
any parallel dimension satisfying $L_{\parallel } /L_{\perp } \gg
1$. Uniform Maxwellian equilibria for ions (protons) and electrons are
chosen, and the correct mass ratio $m_i/m_e=1836$ is used. Spatial
dimensions $(x,y)$ perpendicular to the mean field are treated
pseudospectrally; an upwind finite-difference scheme is used in the
parallel direction, $z$. Collisions are incorporated using a fully
conservative, linearized collision operator that includes energy
diffusion and pitch-angle scattering.\cite{Abel:2008,Barnes:2009}

The perpendicular variation in the simulation is described by a
complex Fourier representation in perpendicular wavevector space
$(k_x,k_y)$. The complex Fourier coefficients must satisfy the reality
condition, for example, $A_\parallel(k_x,k_y) =
A_\parallel^*(-k_x,-k_y)$. Thus, it is necessary to evolve numerically
only the Fourier coefficients in the upper half-plane, $k_y\ge 0$.
Note that on the line defined by $k_y=0$, the Fourier coefficients for
$k_x <0$ are simply the complex conjugates of the Fourier coefficients
on the same line with $k_x>0$.  Additionally, the coefficient
$(k_x,k_y) =(0,0)$, corresponding to a DC offset, is absorbed into the
background equilibrium conditions and is not evolved. We may therefore
restrict our discussion of the nonlinear energy transfer in the
perpendicular plane to the energy associated with the complex Fourier
coefficients in the upper half-plane, $k_y\ge 0$, with the implicit
assumption that coefficients in the lower half-plane are determined by
the reality condition.

The model problem solved in Paper I involves the nonlinear interaction
between two initially overlapping, perpendicularly polarized,
counterpropagating linear \Alfven waves in a periodic domain.  To
simulate this model problem in the \T{AstroGK} code requires the
capability to initialize the linear kinetic eigenfunction for the
\Alfven wave mode, specifying throughout the simulation domain both
the electromagnetic potentials $\phi$, $A_\parallel$, and $\delta
B_\parallel$ and the perturbed gyrokinetic distribution functions
$h_i$ and $h_e$. A specialized module has been written for the
\T{AstroGK} code to accomplish this nontrivial initialization, as described 
in the Appendix.

\section{Numerical Solution}
\label{sec:numsol}
In this section, we present the \T{AstroGK} numerical solution of the
nonlinear interaction between two initially overlapping,
perpendicularly polarized, counterpropagating linear \Alfven waves in
a periodic domain. We consider a uniform, fully ionized proton and
electron plasma with Maxwellian equilibrium distribution functions, a
realistic mass ratio $m_i/m_e=1836$, and a straight, uniform magnetic
field $\V{B}_0 = B_0 \zhat$. The plasma parameters are ion plasma beta
$\beta_i=8 \pi n_i T_i/B_0^2=1$ and ion-to-electron temperature ratio
$T_i/T_e=1$. 

To compare with the analytical incompressible MHD solution in Paper I,
we must choose a simulation domain suitable for investigating the
dynamics in both the anisotropic limit, $k_\perp\gg k_\parallel $, and
the MHD limit, $k_\perp \rho_i \ll 1$.  The simulation domain
$L_{\perp }^2 \times L_{\parallel }$ is therefore taken to have a
perpendicular width $L_\perp = 80 \pi \rho_i$ and is elongated in the
direction of the magnetic field such that the gyrokinetic expansion
parameter $\epsilon = L_\perp/L_\parallel \ll 1$. Here the ion Larmor
radius is defined by $\rho_i = v_{ti}/\Omega_i$, where the ion thermal
velocity is $v_{ti}^2= 2 T_i/m_i$, the ion cyclotron frequency is
$\Omega_i = q_i B_0/(m_i c)$, and the Boltzmann constant has been
absorbed to give temperature in units of energy. Defining the
wavenumbers associated with the domain scale to be $k_\parallel \equiv
2 \pi /L_\parallel$ and $k_\perp\equiv 2 \pi /L_\perp$, the two
initial counterpropagating \Alfven waves will have $k_\perp \rho_i
=0.025$, satisfying the MHD limit, and
$k_\parallel/k_\perp= \epsilon \ll 1$, satisfying the
anisotropic limit.  The characteristic frequency $\omega_0$ of a
domain-scale \Alfven wave is given by $\omega_0 = k_\parallel v_A$.
The numerical solution to the linear collisionless gyrokinetic
dispersion relation gives a real frequency $\omega_r=1.00131 \omega_0$
and damping rate $\gamma=2.94692 \times 10^{-5} \omega_0$.

The dimensions of the \T{AstroGK} numerical simulation are
$(n_x,n_y,n_z,n_\lambda,n_\varepsilon,n_s)= (16,16,32,32,32,2)$. The
collision frequencies used in the linear relaxation phase (see
Appendix) are $\nu_i=2.3 \times 10^{-4} \omega_0$ and $\nu_i=6.5
\times 10^{-5} \omega_0$, and in the nonlinear simulation are 
$\nu_i=1.1 \times 10^{-5} \omega_0$ and $\nu_i=3.2 \times 10^{-6}
\omega_0$. In the weakly damped MHD regime, the linear relaxation phase 
requires many periods to eliminate the transient behavior; in this
simulation at $k_\perp \rho_i =0.025$, this phase continues for 150
periods of the initialized \Alfven waves.  After relaxation, the
linear frequency and damping rate of both of these modes are verified
to give values in agreement with the linear collisionless gyrokinetic
dispersion relation.

In this simulation, we initialize two overlapping, perpendicularly
polarized, counterpropagating linear \Alfven waves, as shown in
Figure~1 of Paper I. Given $k_\perp \rho_i =0.025$ and
$k_\parallel/k_\perp \ll 1$, we specify the initial plane wave modes
by the shorthand $(k_x/k_\perp, k_y/k_\perp,k_z/k_\parallel)=(1,0,-1)$
and $(0,1,1)$. Note that we maintain the convention established in
Paper I that $\omega_0 = k_\parallel v_A >0 $, so that the sign of
$k_z$ determines the direction of propagation of the \Alfven wave;
therefore, it is clear that the initialized waves are both
perpendicularly polarized and counterpropagating.

A critical aspect of the nonlinear simulation is to select the
amplitude of the initialized modes to satisfy the ordering assumed in
the asymptotic analytical solution in Paper I. For the incompressible
MHD equations in the symmetrized Els\"asser form,\citep{Elsasser:1950}
\begin{equation}
\frac{\partial \V{z}^{\pm}}{\partial t} 
\mp \V{v}_A \cdot \nabla \V{z}^{\pm} 
=-  \V{z}^{\mp}\cdot \nabla \V{z}^{\pm} -\nabla P/\rho_0,
\label{eq:elsasserpm}
\end{equation}
this ordering imposes that magnitude of the nonlinear term
$\V{z}^{\mp}\cdot \nabla \V{z}^{\pm}$ is small compared to the
magnitude of the linear term $\V{v}_A \cdot \nabla \V{z}^{\pm}$.  The
strength of the nonlinearity affecting the evolution of $ \V{z}^{\pm}$
is conveniently quantified by defining a nonlinearity parameter $\chi ^\pm$ that
is the ratio of the magnitude of the nonlinear term to that of the
linear term
\begin{equation}
\chi ^\pm \equiv \left|\frac{\V{k}_\perp^\pm \cdot \V{z}^\mp}{k_\parallel^\pm v_A}\right|
\label{eq:chi}
\end{equation}
A nonlinearity parameter $\chi ^\pm \sim 1$ signifies the limit of
critically balanced, strong turbulence,\cite{Goldreich:1995} where the
linear and nonlinear terms have the same magnitude; the limit $\chi
^\pm \ll 1$ corresponds to the case of weak turbulence, and is the
appropriate case to compare to the analytical solution.  Substituting
the definition of the Elsasser potentials, $\V{z}^{\mp} =
\zhat \times \nabla_\perp \zeta^{\mp}$, into \eqref{eq:chi}, we define a 
characteristic amplitude $\zeta_{NL}^\mp$ that corresponds to the case of 
critically balanced, strong turbulence with $\chi ^\pm =1$, given by 
\begin{equation}
\zeta_{NL}^\mp \equiv \frac{k_\parallel^\pm v_A}{\zhat \cdot(\V{k}_\perp^\mp \times \V{k}_\perp^\pm)}.
\end{equation}
With this definition, the nonlinearity parameter for a
counterpropagating \Alfven wave collision is simply given by $\chi
^\pm =\zeta^\mp/\zeta_{NL}^\mp$. For the two equal-amplitude,
counterpropagating \Alfven waves specified in this simulation, the
characteristic amplitudes are the same for both wave directions $\zeta_{NL}^+=\zeta_{NL}^-$,
reducing the expression above to $\zeta_{NL} \equiv k_\parallel v_A/k_\perp^2$. The
amplitudes of the initial waves for the simulation are specified to be
$\zeta^\mp/\zeta_{NL}=0.02$, satisfying the ordering assumed in the
analytical calculation.

In closing, we state for completeness that the relationship between
the Elsasser potentials and the gyrokinetic
potentials\cite{Schekochihin:2009} is given by
\begin{equation}
\zeta^\pm = \frac{c \phi}{B_0} \mp \frac{A_\parallel}{\sqrt{4 \pi n_0 m_i}}.
\end{equation}

\subsection{Overall Nonlinear Evolution}
\begin{figure}
\hbox to \hsize{ \hfill 
\resizebox{3.3in}{!}{\includegraphics{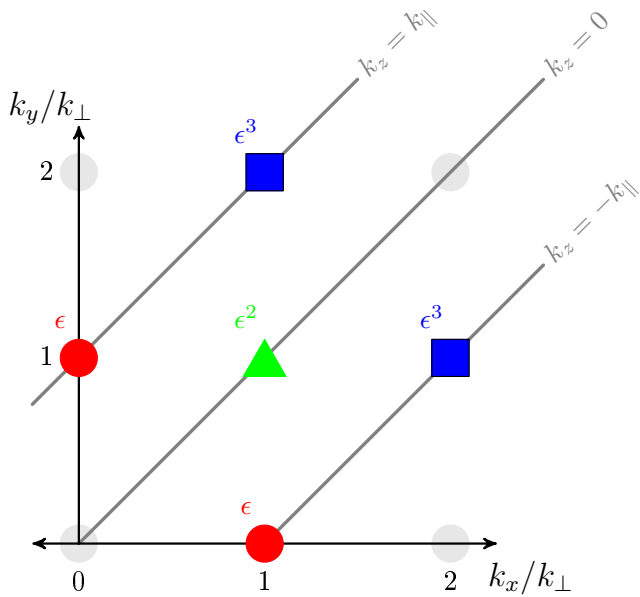}}\hfill}
\caption{ \label{fig:modes}  Schematic diagram of the Fourier modes that 
play a role in the secular transfer of energy to small scales in an
\Alfven wave collision. These key Fourier modes are the  primary counterpropagating
\Alfven waves (red circles), the secondary inherently
nonlinear magnetic field fluctuation (green triangle), and the tertiary
counterpropagating \Alfven waves (blue squares). The parallel wavenumber $k_z$
for each of the modes is indicated by the diagonal grey lines, a
consequence of the resonance conditions for the wavevector. }
\end{figure}

We begin the presentation of the \T{AstroGK} numerical solution with
an overall picture of the energy transfer due to the nonlinear
interaction between two counterpropagating \Alfven waves.  In this
section, we focus only on the Fourier modes that play a role in the
secular transfer of energy to smaller scales, or larger wavenumbers,
as described qualitatively in the discussion in \S IV A of Paper I.
The key modes at each asymptotic order of the analytical solution,
presented in \figref{fig:modes}, are the primary counterpropagating
\Alfven waves $(1,0,-1)$ and $(0,1,1)$ (red circles), the secondary inherently
nonlinear magnetic field fluctuation $(1,1,0)$ (green triangle), and the
tertiary counterpropagating \Alfven waves $(2,1,-1)$ and $(1,2,1)$
(blue squares). 

\begin{figure}
\hbox to \hsize{ \hfill 
\resizebox{3.3in}{!}{\includegraphics{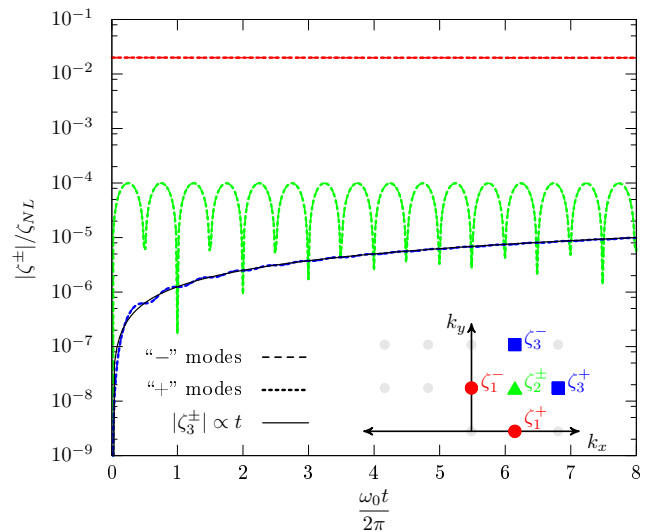}}\hfill}
\caption{ \label{fig:evol}  Evolution of the normalized amplitude 
$|\zeta^\pm|/\zeta_{NL}$ of the key Fourier modes vs.~time $\omega_0 t
/ 2 \pi$ over eight periods of the primary \Alfven waves. The color 
map is the same as \figref{fig:modes}, and a linear increase with time
is indicated by the solid black line. }
\end{figure}

The time evolution of the normalized amplitudes of
$|\zeta^\pm|/\zeta_{NL}$ for each of these key Fourier modes is shown
in \figref{fig:evol}, demonstrating a number of important qualitative
characteristics of the nonlinear evolution.  First, note that
solutions remain well-ordered throughout the evolution of the
simulation, with $|\zeta_1^\pm| \gg |\zeta_2^\pm| \gg
|\zeta_3^\pm|$. Therefore, we expect that the asymptotic analytical
solution derived Paper I should remain valid over the entire
simulation. Second, the due to the weak nonlinearity
$\zeta^\pm/\zeta_{NL}=0.02$, the energy loss from the primary modes
(red) is negligible.  Third, the second-order mode (green) indeed has
a frequency of $2 \omega_0$, as expected from the analytical solution.
Finally, the secular energy gain by the tertiary modes (blue) leads to
an amplitude that increases linearly with time (solid black),
$|\zeta_3^\pm|\propto t$, as predicted by the analytical solution.

\subsection{Numerical Validation of Analytical Solution to $\Order(\epsilon^3)$  }
\label{sec:valid}
\begin{figure*}
\hbox to \hsize{ \hfill 
\resizebox{6.6in}{!}{\includegraphics{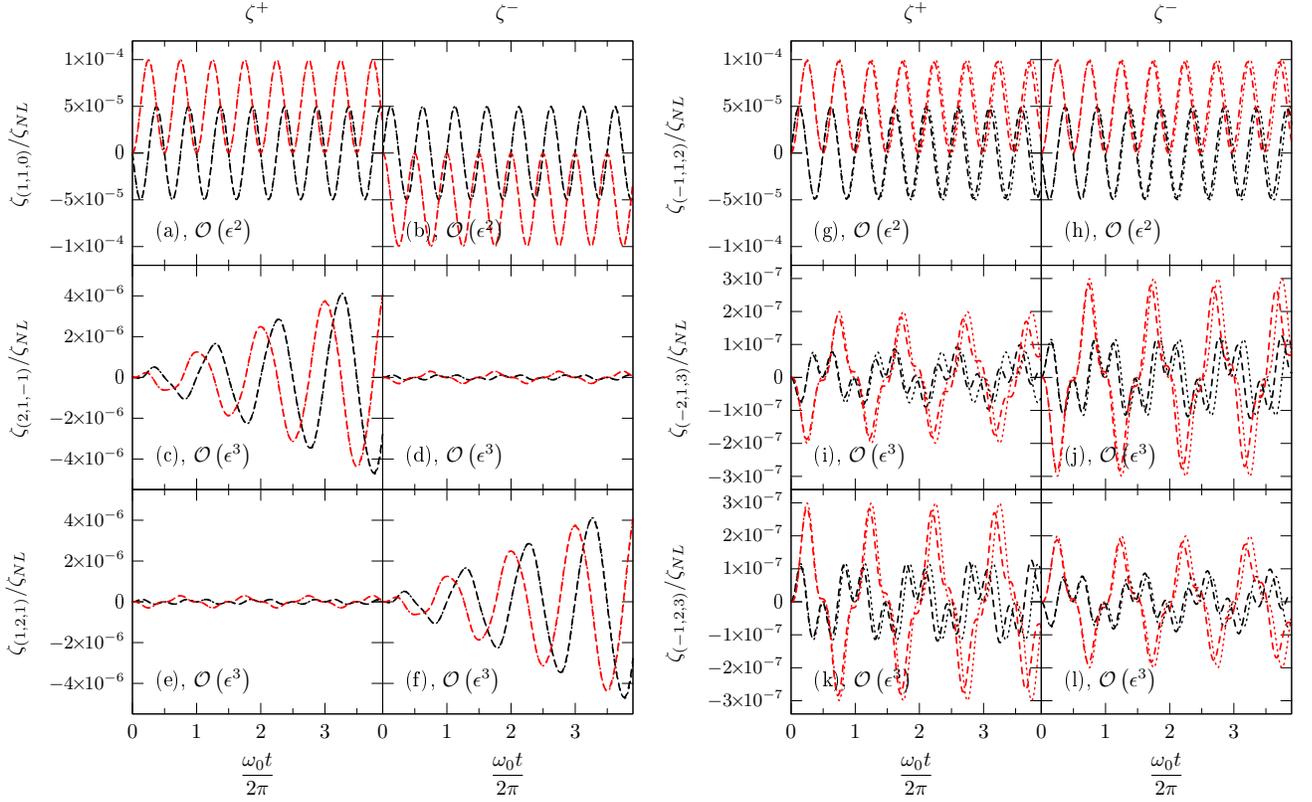}}\hfill}
\caption{ \label{fig:all} A comparison of the  real (black) and imaginary (red) components of the complex
Elsasser potentials $\zeta_n^\pm$ at orders $n=1,2,3$ from the
asymptotic analytical solution (dotted) and gyrokinetic numerical
simulation (dashed).}
\end{figure*}

Here we present a thorough validation of the asymptotic analytical
solution in Paper I for all of the modes arising up to
$\Order(\epsilon^3)$.  All of these modes are depicted in Figure~2 of
Paper I, which shows both the key Fourier modes shown here in
\figref{fig:modes} and the Fourier modes at $k_x <0$ that do not play
a role in the secular energy transfer. In \figref{fig:all}, we plot
the real (black) and imaginary (red) components of the complex
Elsasser potentials $\zeta_n^\pm$ at orders $n=1,2,3$ from the
asymptotic analytical solution (dotted) and gyrokinetic numerical
simulation (dashed). The left panel presents the modes with $k_x >0$
that play a role in the secular energy transfer to small scales, and
the right panel presents the modes with $k_x<0$ that do not play a
role in this energy transfer. All modes with $k_x >0$ (panels a--f)
show excellent agreement between the asymptotic analytical solution
and the gyrokinetic numerical simulation. The modes with $k_x <0$
(panels g--l) show that a phase difference arises over time between
the analytical and numerical solutions.  The cause of this minor
discrepancy is not clear, but may be due to higher-order effects not
included in the incompressible MHD solution arising from finite Larmor
radius corrections.  Note, however, the amplitude of the third order
solutions in \figref{fig:all} (lower two rows across both panels)
demonstrates that the modes in panels (c) and (f) dominate the energy
in the tertiary solutions, and both of these modes show excellent
agreement with the analytical predictions.  Therefore, the minor phase
differences that arise for the $k_x <0$ modes do not suggest an
energetically significant deviation from the analytical solution.

\subsection{Physical Representation of the Solution in $\V{B}$ and $\V{E}$ }
Although the comparison of the nonlinear evolution of the analytical
and numerical complex Elsasser potentials in \secref{sec:valid}
provides a thorough validation of the analytical solution, it does not
immediately provide a simple intuitive picture of the dynamical
evolution of the turbulent electromagnetic fields. Since it is the
real electromagnetic fields that are measured directly
in a turbulent plasma, examining this evolution is of vital
importance.  Therefore, we present here a complementary comparison
between the analytical and numerical solutions of the evolution of the
magnetic and electric field fluctuations. In Paper I, the secondary $\Order(\epsilon^2)$
solutions for $\V{B}_{\perp 2}$ and $\V{E}_{\perp 2}$ are given by
(36) and (37), and the tertiary $\Order(\epsilon^3)$ solutions for $\V{B}_{\perp 3}$ and
$\V{E}_{\perp 3}$ are given by (40) and (41).

\begin{figure}
\resizebox{3.3in}{!}{\includegraphics{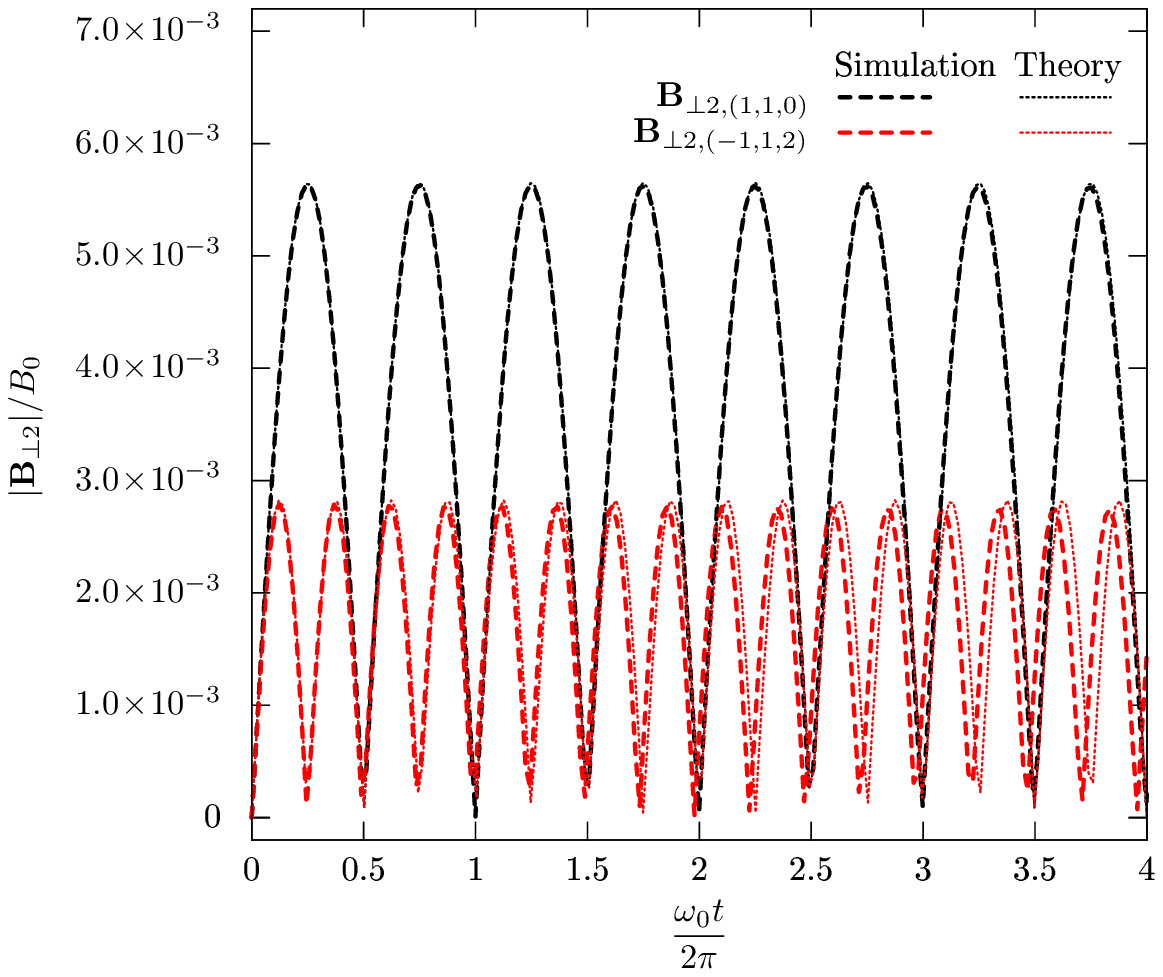}}
\resizebox{3.3in}{!}{\includegraphics{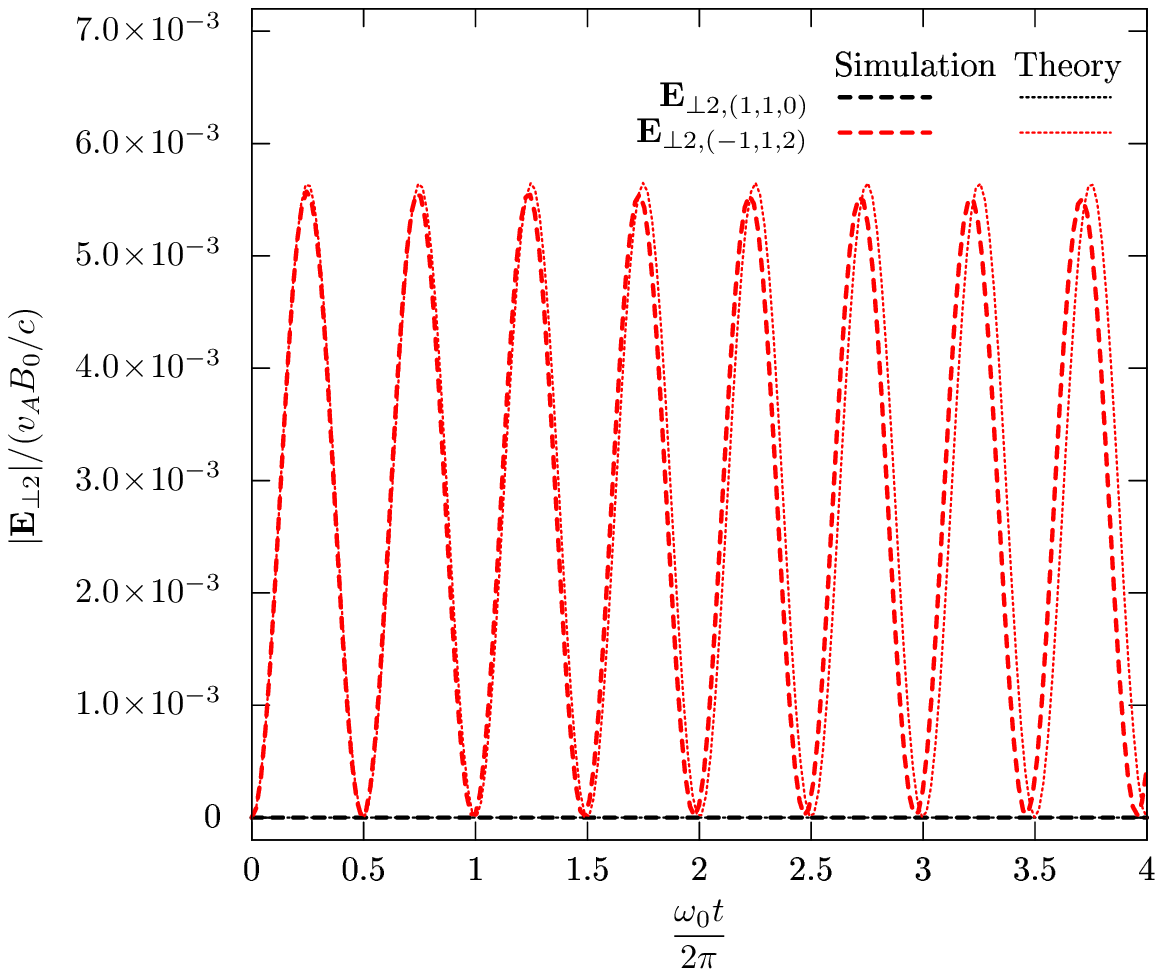}}
\caption{ \label{fig:wt_O2} Comparison of the analytical (dotted) and  numerical
(dashed) solutions for the time evolution of the amplitude of the
magnetic and electric fields at $\Order(\epsilon^2)$. The top panel
shows the normalized amplitude of the magnetic field $|\V{B}_{\perp
2}|/B_0$ for the $(1,1,0)$ mode (black) and $(-1,1,2)$ (red). The
bottom panel shows the normalized amplitude of the electric field
$|\V{E}_{\perp 2}|/(v_A B_0/c)$ for the $(1,1,0)$ mode (black) and
$(-1,1,2)$ mode (red). }
\end{figure}

The second order $\Order(\epsilon^2)$ asymptotic solution consists of
two Fourier modes $(1,1,0)$ and $(-1,1,2)$ according to (36) and (37)
of Paper I. The normalized amplitude of the magnetic field
$|\V{B}_{\perp 2}|/B_0$ is plotted in the top panel of
\figref{fig:wt_O2} for the  $(1,1,0)$ mode (black) and $(-1,1,2)$ mode (red) 
from both the analytical calculation (dotted) and the numerical
solution (dashed). Similarly, the normalized amplitude of the electric
field $|\V{E}_{\perp 2}|/(v_A B_0/c)$ is plotted in the bottom panel of
\figref{fig:wt_O2} for the  $(1,1,0)$ mode (black) and $(-1,1,2)$ mode (red) 
from both the analytical calculation (dotted) and the numerical
solution (dashed). The $\Order(\epsilon^2)$ nonlinear response in
\figref{fig:wt_O2} is purely oscillatory, with no secular energy
transfer of energy at this order.  The $(1,1,0)$ mode solutions agree
excellently, with only a magnetic field response and no corresponding
electric field response---this is the inherently nonlinear $k_z=0$
magnetic fluctuation at $\Order(\epsilon^2)$ that plays the crucial
role in the secular energy transfer to the tertiary
$\Order(\epsilon^3)$ \Alfven waves, as emphasized in Paper I. The
$(-1,1,2)$ modes also agree very closely initially, but eventually a
small phase shift arises between the numerical and analytical
solution, similar to that seen in
\figref{fig:all}.

Using trigonometric addition formulas, the form of the time dependence
of the amplitude of the $\Order(\epsilon^2)$ modes can be written as
\begin{equation}
|\V{B}_{\perp 2, (1,1,0)}| \propto \sqrt{1-\cos(2 \omega_0t)},
\end{equation}
\begin{equation}
|\V{B}_{\perp 2, (-1,1,2)}| \propto |\sin(2 \omega_0t)|,
\end{equation}
\begin{equation}
|\V{E}_{\perp 2, (1,1,0)}| =0,
\end{equation}
\begin{equation}
|\V{E}_{\perp 2, (-1,1,2)}| \propto  1-\cos(2 \omega_0t).
\end{equation}

The third order $\Order(\epsilon^3)$ of the asymptotic solution in
Paper I is given by (40) and (41) and consists of two Fourier modes
$(2,1,-1)$ and $(1,2,1)$ whose amplitudes increase secularly, and four
Fourier modes that display oscillatory behavior, $(-2,1,3)$,
$(-1,2,3)$, $(0,1,1)$, and $(1,0,-1)$. Below we present a comparison
between the analytical and numerical solutions for the first four
Fourier modes; the latter two modes, $(0,1,1)$ and $(1,0,-1)$, merely
represent $\Order(\epsilon^3)$ corrections to the primary wave
amplitudes and will not be considered further.

\begin{figure}
\resizebox{3.3in}{!}{\includegraphics{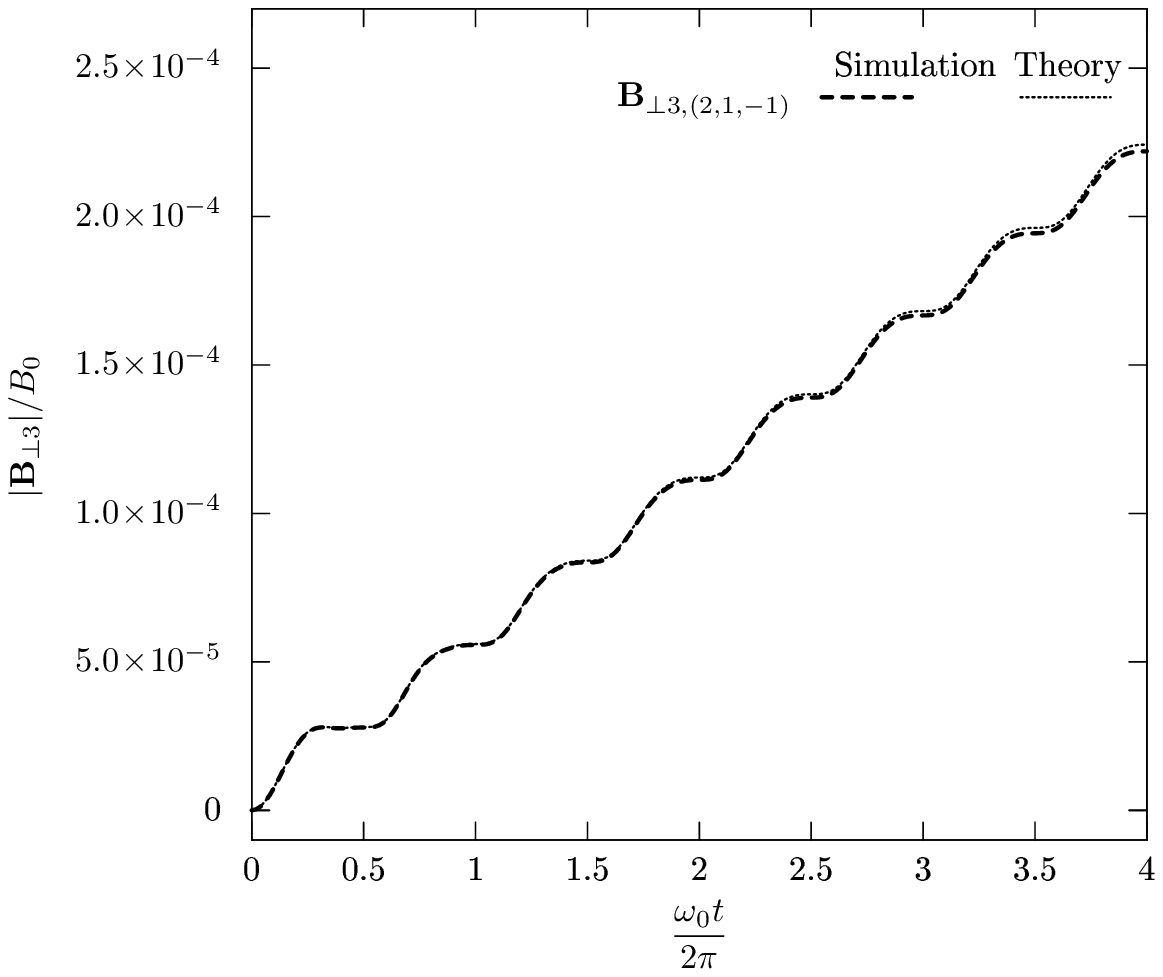}}
\resizebox{3.3in}{!}{\includegraphics{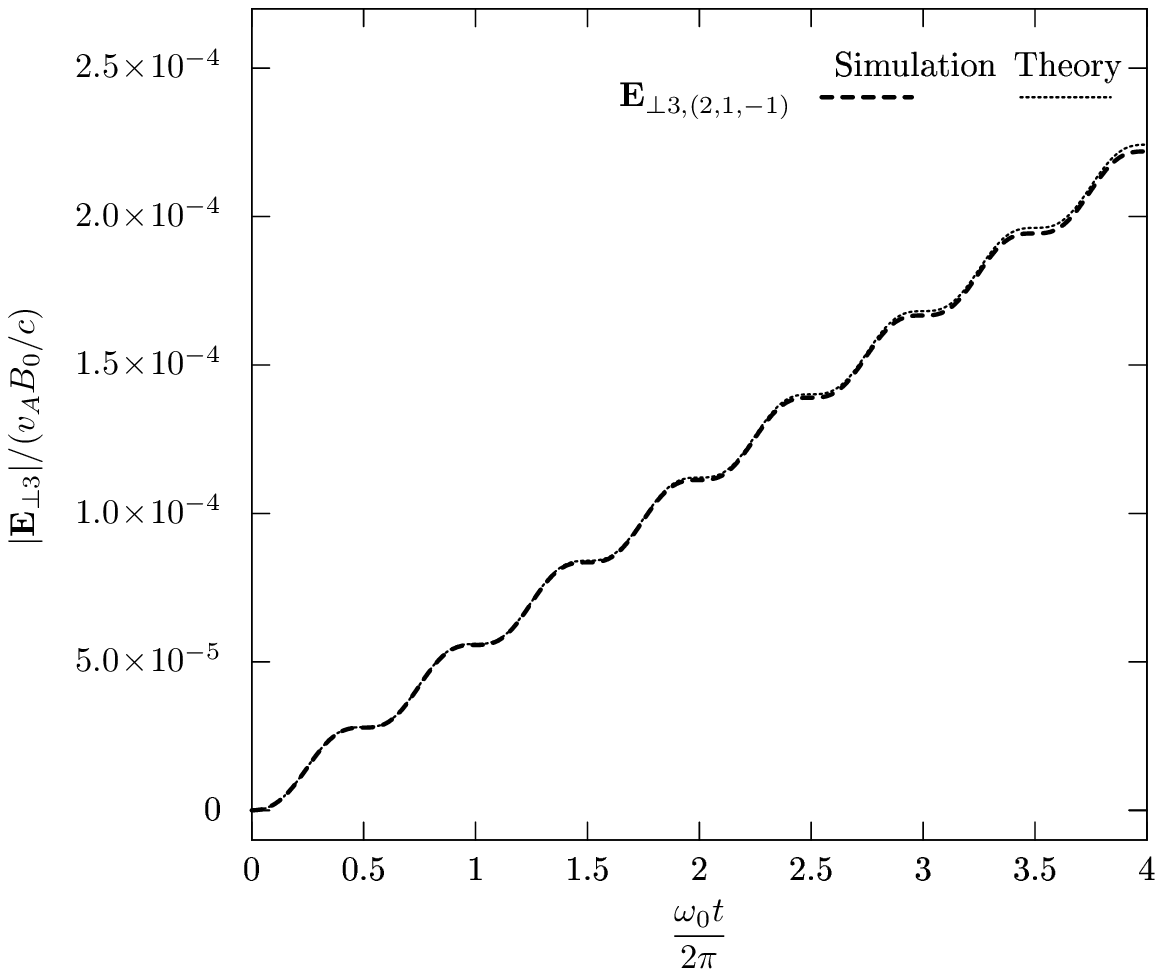}}
\caption{Comparison of the analytical (dotted) and  numerical
(dashed) solutions for the time evolution of the amplitude of the
magnetic (top) and electric (bottom) fields of the $(2,1,-1)$ Fourier
mode at $\Order(\epsilon^3)$.
}
\label{fig:wt_O3_2x1y}
\end{figure}

The energetically dominant $\Order(\epsilon^3)$ Fourier modes are the
two modes, $(2,1,-1)$ and $(1,2,1)$, that receive a secular transfer
of energy through the nonlinear interaction.  As discussed in Paper I,
these two modes are \Alfven waves with the same value of $k_z$ as the
two primary \Alfven waves, indicating no parallel cascade of energy,
but a higher value of the perpendicular component of the wavenumber.
The transfer of energy to these tertiary $\Order(\epsilon^3)$
\Alfven waves represents the nonlinear cascade of energy to smaller 
scales that is the most important effect of turbulence in
astrophysical plasmas. In \figref{fig:wt_O3_2x1y}, we plot the
analytical (dotted) and numerical (dashed) solutions of the normalized
amplitude of the magnetic field $|\V{B}_{\perp 2}|/B_0$ (top panel)
and electric field $|\V{E}_{\perp 2}|/(v_A B_0/c)$ (bottom panel) for
the $(2,1,-1)$ mode. The agreement between the analytical and
numerical solution is seen to be excellent for both fields. Note that,
by the inspection of (40) and (41) from Paper I, it is clear that the
evolution of the $(1,2,1)$ mode has the same form as that for the
$(2,1,-1)$ mode, only with the electric and magnetic fields swapped;
therefore, we do not provide a separate plot for the $(1,2,1)$ mode,
but note that the agreement is the same as that shown in
\figref{fig:wt_O3_2x1y}. 

An important dynamical feature of the nonlinear energy transfer to
small scales is evident in \figref{fig:wt_O3_2x1y}.  As discussed in
Paper I, the purely magnetic $(1,1,0)$ mode of the
$\Order(\epsilon^2)$ solution plays a crucial role in the secular
energy transfer from the primary $(0,1,1)$ and $(1,0,-1)$ \Alfven
waves to the tertiary $(2,1,-1)$ and $(1,2,1)$ \Alfven waves.  It is
evident, upon examination of the time evolution of the $(1,1,0)$ mode
in \figref{fig:wt_O2} and of the $(2,1,-1)$ mode in
\figref{fig:wt_O3_2x1y}, that the amplitude of the $(2,1,-1)$ mode
increases only at times when the $(1,1,0)$ mode has non-zero
amplitude. Conversely, at times $\omega_0 t/2 \pi = n/2$ for $n=0,1,2,
\ldots$, when the amplitude of the secondary $(1,1,0)$ mode is zero,
the amplitude of the $(2,1,-1)$ mode does not increase.  This feature
of the solution highlights the essential role played by the inherently
nonlinear, purely magnetic, secondary $(1,1,0)$ mode in the secular
transfer of energy to small scales.


\begin{figure}
\resizebox{3.3in}{!}{\includegraphics{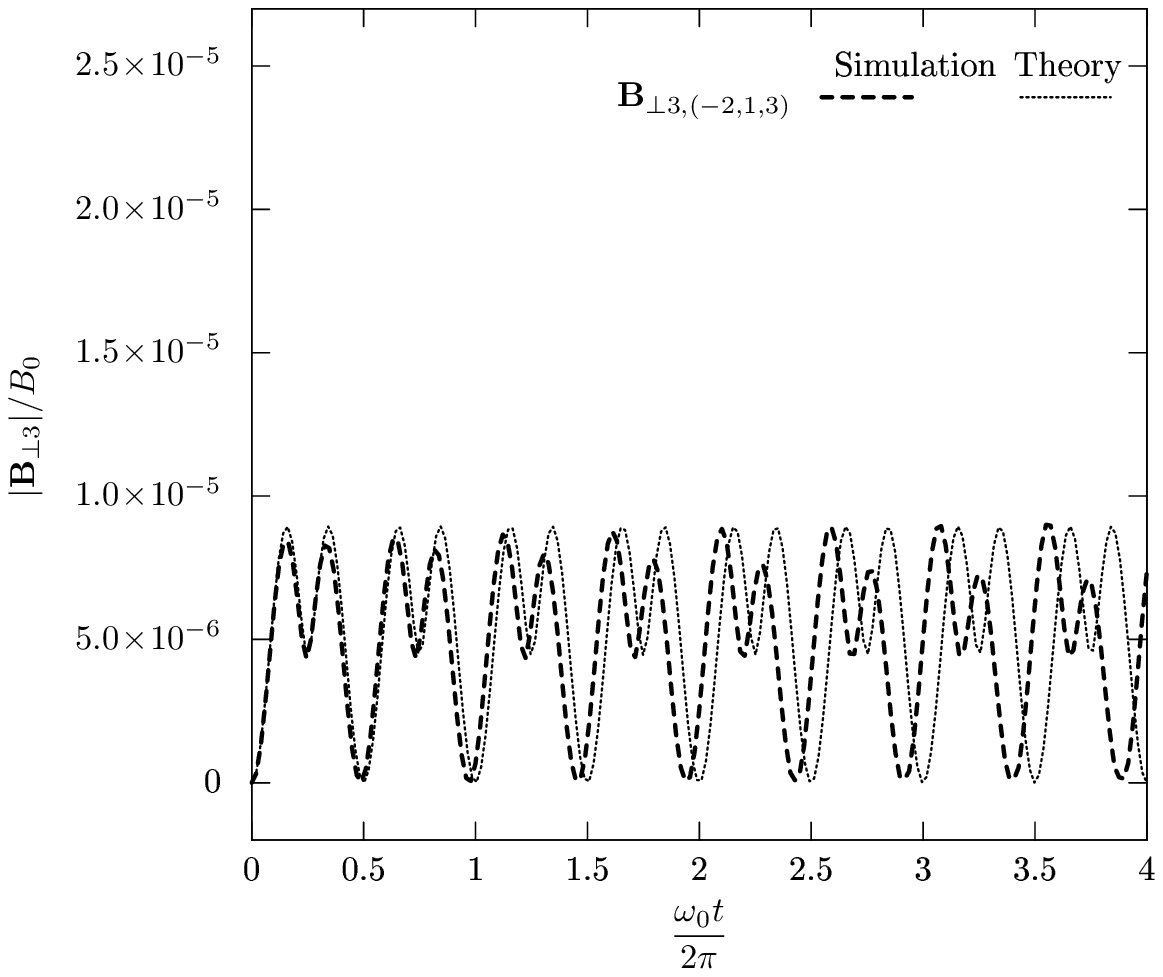}}
\resizebox{3.3in}{!}{\includegraphics{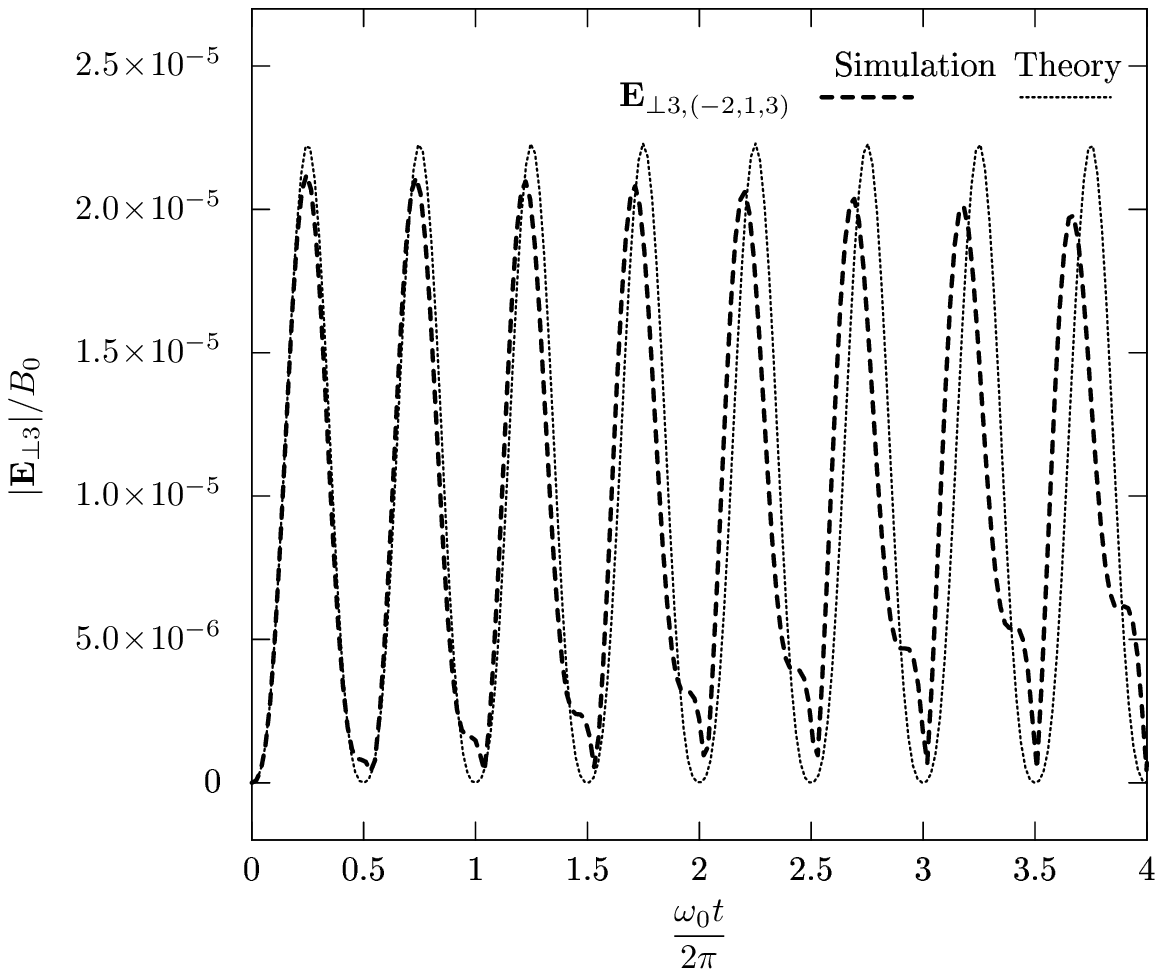}}
\caption{Comparison of the analytical (dotted) and  numerical
(dashed) solutions for the time evolution of the amplitude of the
magnetic (top) and electric (bottom) fields of the $(-2,1,3)$ Fourier
mode at $\Order(\epsilon^3)$.
}
\label{fig:wt_O3_m2x1y}
\end{figure}

To complete our comparison of the analytical and numerical solutions
for the magnetic and electric field evolution, we turn our attention
to the nonlinear response of the $(-2,1,3)$ and $(-1,2,3)$ Fourier
modes in the $\Order(\epsilon^3)$ solution. In
\figref{fig:wt_O3_m2x1y}, we plot the analytical (dotted) and
numerical (dashed) solutions of the normalized amplitude of the
magnetic field $|\V{B}_{\perp 2}|/B_0$ (top panel) and electric field
$|\V{E}_{\perp 2}|/(v_A B_0/c)$ (bottom panel) for the $(-2,1,3)$
mode. Although agreement with the analytical solution at early times
is excellent, a phase shift and alteration of the form of the solution
arises after several primary wave periods. Note that the solutions for
the magnetic and electric field evolution of the $(-1,2,3)$ Fourier
mode is the same as the $(-2,1,3)$ mode presented in
\figref{fig:wt_O3_m2x1y}.


\section{Discussion and Conclusion}
\label{sec:discuss}
In Paper I, we have derived an asymptotic analytical solution for the
nonlinear interaction between two initially overlapping,
perpendicularly polarized, counterpropagating linear \Alfven waves in
the weakly nonlinear limit.  The incompressible MHD solution to this
idealized problem provides valuable intuition into the fundamental
nature of the nonlinear transfer of energy to small scales in
magnetized plasma turbulence. The incompressible MHD solution in Paper
I is formally rigorous in the anisotropic limit, $k_\perp \gg
k_\parallel$; in this anisotropic limit, the incompressible MHD
solution is equivalent to a reduced MHD solution for the \Alfvenic
dynamics. The primary aim of this companion paper is to present a
thorough numerical verification of the analytical solution using the
gyrokinetic code \T{AstroGK} in the MHD limit, $k_\perp \rho_i \ll 1$.

The numerical validation shows excellent agreement between the
analytical solution and the nonlinear gyrokinetic simulation results
for the Fourier modes that play a role in the secular transfer of
energy from the primary \Alfven waves to the tertiary \Alfven waves,
as depicted in \figref{fig:modes}. For the Fourier modes with $k_x
<0$, modes that do not play a role in this secular energy transfer,
the agreement is very good at early times, but a small phase shift
arises after several periods of the primary waves. This minor
discrepancy may arise through dispersive effects due to finite Larmor
radius corrections that are included in the gyrokinetic simulation but
not in the incompressible MHD solution.  We note, however, that the
modes suffering this small phase shift are energetically subdominant
to the modes that receive the secular transfer of energy from the
primary \Alfven waves.  In summary, the results presented here verify
the analytical solution derived in Paper I, accomplishing the primary
aim of this paper.

The numerical solution presented here illustrates a couple of salient
features of the nonlinear energy transfer. First, as evident in
\figref{fig:evol}, the lowest order of the solution displaying a secular 
increase of energy is $\Order(\epsilon^3)$. As discussed at length in
Paper I, this is a consequence of the fact that the energy transfer is
due to a resonant, four-wave interaction. Second, also seen in 
\figref{fig:evol}, the secular increase of amplitude of these tertiary 
modes is linear in time, corresponding to an increase in energy
$\propto t^2$. Apparently inconsistent with the expectation from
scaling theories of turbulence, this characteristic scaling follows
from the coherent nature of the interaction between the primary
counterpropagating \Alfven waves.  For a more realistic picture of
plasma turbulence involving the cumulative effect of successive
nonlinear interactions between many uncorrelated \Alfven wave packets,
accounting for a random walk in energy changes the scaling of the
increase in energy to $\propto t $, as expected from turbulence
theories.

The secondary aim of this paper is to test the hypothesis that the
physical mechanism for nonlinear energy transfer in plasma turbulence
under astrophysically relevant conditions remains well described by
the incompressible MHD solution derived in Paper I. In particular,
turbulent astrophysical plasmas are often found both to be weakly
collisional and to have a typical plasma beta $\beta \lesssim 1$, two
limits in which the equations of incompressible MHD are formally
invalid. To test this hypothesis, we compare the analytical
incompressible MHD solution to numerical gyrokinetic solution in the
MHD limit, $k_\perp \rho_i \ll 1$, using the Astrophysical
Gyrokinetics code, \T{AstroGK}. Since the gyrokinetic equations
rigorously describe the low-frequency kinetic dynamics of the
turbulence in the anisotropic limit, the demonstrated agreement
between the two methods signifies that, indeed, the incompressible MHD
solution satisfactorily describes, in a simple analytical form, the
essential dynamics of the nonlinear energy transfer in a turbulent,
weakly collisional plasma.

This result was anticipated by the theoretical
finding\cite{Schekochihin:2009} that the equations of reduced MHD
rigorously describe the kinetic dynamics of the turbulent \Alfvenic
fluctuations in the limit $k_\perp \rho_i
\ll 1$.  Such a surprising simplification of the kinetic dynamics to 
a fluid limit can be understood physically as resulting from the
incompressible nature of \Alfven waves in the limit $k_\perp \rho_i
\ll 1$. Since \Alfven waves have no associated  motions parallel to the 
magnetic field, the plasma collisionality has influence on neither the
linear nor the nonlinear dynamics of \Alfvenic fluctuations.

In conclusion, we have numerically verified the asymptotic analytical
solution in Paper I\cite{Howes:2013a} for the nonlinear interaction between
counterpropagating \Alfven waves in the weakly nonlinear limit.  In
addition, by comparing the analytical incompressible MHD solution to
numerical gyrokinetic solution, we have confirmed the hypothesis that
that the physical mechanism underlying the nonlinear energy transfer
in plasma turbulence under astrophysically relevant conditions remains
well described by the simple fluid description of incompressible MHD.

These findings motivate a simplified picture for describing the very
complex phenomenon of turbulence in a kinetic plasma.  The development
of a thorough understanding of kinetic turbulence, a key goal of the
space physics and astrophysics communities, requires the elucidation
of (1) the fundamental physical mechanisms underlying the nonlinear energy
transfer from large to small scales, (2)  the dissipation of the turbulent
fluctuations at small scales, and (3) the ultimate conversion of the
turbulent energy into plasma heat. Our findings suggest that
illuminating the nonlinear wave-wave interactions responsible for the
turbulent cascade of energy from large to small scales does not
necessarily require a kinetic treatment, but that the dynamics can be
satisfactorily described using a reduced fluid description. In the MHD
limit, $k_\perp \rho_i \ll 1$, we have proven here that the
incompressible MHD solution is sufficient; for the limit of a
turbulent kinetic \Alfven wave cascade, $k_\perp \rho_i \gg 1$, an
appropriate fluid description, such as electron reduced
MHD,\cite{Schekochihin:2009} may be sufficient to describe the
nonlinear wave-wave interactions underlying the turbulent energy
transfer in this regime. On the other hand, the physical mechanisms in kinetic
turbulence responsible for the dissipation and thermalization of the
turbulent energy almost certainly do require a kinetic description.
The closures for dissipation typically used in fluid descriptions,
such as viscosity and resistivity, are not valid in the weakly
collisional limit relevant to turbulent dissipation in space and
astrophysical plasmas.  Instead, inherently kinetic mechanisms, such
as collisionless wave-particle interactions and infrequent particle
collisions, are almost certainly responsible for the damping of the
turbulent fluctuations and the ultimate conversion of their energy into
plasma
heat.\cite{Howes:2006,Howes:2008b,Schekochihin:2009,Howes:2011a,Howes:2011b,TenBarge:2012d}.

Based on these arguments, we propose the following simplified
framework for understanding kinetic plasma turbulence: (a) nonlinear
wave-wave interactions are responsible for the turbulent cascade of
energy from large to small scales, and can be adequately described
using an appropriate fluid description; (b) kinetic mechanisms, such
as collisionless wave-particle interactions, are responsible for the
damping of the turbulent electromagnetic fluctuations, requiring a
kinetic description; and, (c) thermalization of the free energy in the
particle distribution functions, a consequence of  the wave-particle
interactions above, requires particle collisions to increase entropy and
realize irreversible thermodynamic heating,\cite{Howes:2006} a process
mediated by an inherently kinetic entropy
cascade.\cite{Schekochihin:2009,Tatsuno:2009}

The findings presented here and in Paper I present a simple picture of
the nonlinear energy transfer in \Alfven wave collisions, establish
the validity of the derived incompressible MHD solution, and
demonstrate its relevance to turbulence in astrophysical plasmas in
the weakly nonlinear limit.  These analytical and numerical solutions
have played an invaluable role in the design of an experiment to
measure the nonlinear interaction between counterpropagating \Alfven
waves in the laboratory.\cite{Howes:2012b} Future extensions of this
work will determine whether the essential characteristics of the
nonlinear energy transfer persist into the important regime of strong
turbulence. In addition, we aim to explore how the nature of the
nonlinear energy transfer changes as the turbulent cascade to small
scales enters the dispersive regime of kinetic \Alfven waves.

\begin{acknowledgments}
This work was supported by NSF PHY-10033446, NSF CAREER AGS-1054061,
and NASA NNX10AC91G. Computing resources were supplied through DOE
INCITE Award PSS002, NSF TeraGrid Award PHY090084, and DOE INCITE
Award FUS030.
\end{acknowledgments}

%

\appendix
\section{Eigenfunction Initialization and Transient Elimination}
In this appendix, we describe the procedure for initialization of
linear gyrokinetic eigenfunctions and for the elimination of
transient behavior in the initialized modes.

First, the plane wave mode to be initialized is chosen by specifying
its wavevector $\V{k}= k_x \xhat + k_y \yhat + k_z \zhat$ and
providing an initial guess for the complex frequency $\omega= \omega_r
-i \gamma$ of the mode.  This guess for the frequency is necessary to
ensure that the correct linear wave mode is initialized---in the MHD
limit of gyrokinetics, both \Alfven waves and kinetic slow waves
\cite{Klein:2012} are possible, as well as a non-propagating entropy
mode.  For the problem at hand, we specify that both initialized modes
are \Alfven waves. Second, the code employs a numerical solver for the
linear, collisionless, gyrokinetic dispersion
relation\cite{Howes:2006} to solve for the complex eigenfrequency
$\omega$ and the complex Fourier coefficients for the eigenfunctions
of the electromagnetic potentials $\hat{\phi}$, $\hat{A}_\parallel$
and $\delta \hat{B}_\parallel$, where the hat symbol denotes the
Fourier coefficient.  Third, these coefficients are used to initialize
at $t=0$ the electromagnetic potentials on the numerical grid
according to
$\hat{A}_\parallel(k_x,k_y,z,t)=\hat{A}_\parallel(k_x,k_y) \exp(i k_z
z - i\omega t + i\delta)$, where $\delta$ allows for an arbitrary
adjustment to the phase of each initialized wave. Fourth, these initial values of
the electromagnetic potentials and the complex eigenfrequency are used
to compute the complex Fourier coefficients in the perpendicular plane
of the perturbed gyrokinetic distribution functions $\hat{h}_i$ and
$\hat{h}_e$, according the eq~(C6) in Howes \emph{et
al.}\cite{Howes:2006} Note that these five-dimensional gyrokinetic
distribution functions are functions of not only the three spatial
coordinates $(k_x,k_y,z)$ but also the two coordinates of gyroaveraged
velocity space, $v_\perp$ and $v_\parallel$. The generic functional
form is therefore
$\hat{h}_s(k_x,k_y,z,\lambda,\varepsilon)=
\hat{h}_s\left[\omega,\hat{\phi}(k_x,k_y,z),\hat{A}_\parallel(k_x,k_y,z),\delta
\hat{B}_\parallel(k_x,k_y,z),v_\perp,v_\parallel\right]$.
This completes the initialization of a single plane wave mode;
additional modes to be initialized may be added by linear
superposition.

Thorough testing of this exact linear eigenfunction initialization
module has exposed a transient behavior of the initialized mode that
does not follow the expectations from the linear theory.  There exist
two potential causes for this transient behavior.  First, the
eigenfunction is computed by the numerical solution of an analytical
form for the linear dispersion relation which assumes a continuous
representation of the distribution functions in velocity space, yet
the numerical representation of velocity space in \T{AstroGK} is
discrete. Therefore, transient behavior that does not follow the
linear analytical theory may arise from slight differences between the
eigenfunctions in discrete and continuous velocity-space
representations. Second, the linear dispersion relation that is solved
is collisionless, yet a non-zero collisionality is always employed in
\T{AstroGK} to maintain resolved structures in velocity
space,\cite{Barnes:2010,Howes:2011a} and finite collisionality may
lead to deviations of the weakly collisional eigenfunctions in the
numerical code from the collisionless eigenfunctions in the analytical
theory.  Whatever the cause of this transient behavior, a simple
procedure has been developed that quickly and effectively allows the
initialized modes to relax to behavior consistent with the linear
theory, as detailed below.

To eliminate the transient behavior, it suffices to perform a linear
relaxation of the initialized mode by running \T{AstroGK} in linear
mode (with the nonlinear terms turned off) with an enhanced
collisionality for a number of wave periods sufficient to eliminate
this transient behavior.  This enhanced collisionality effectively
eliminates the deviations of the initialized eigenfunction from the
eigenfunction that is appropriate for the discrete, weakly collisional
representation in the \T{AstroGK}. Consider a linear collisionless
gyrokinetic wave mode that has real frequency $\omega_r$ and damping
rate $\gamma= \gamma_i + \gamma_e$, where $\gamma_s$ is the
collisionless damping rate due to the Landau resonance with species
$s$. \T{AstroGK} simulations are typically run with collisionalities
set to values of $\nu_s \le 0.5 \gamma_s$.  A value of collisionality of
$\nu_s = 0.5 \gamma_s$ is sufficiently high to ensure that the structure
in velocity space that is generated by wave-particle interactions remains
resolved\cite{Barnes:2010,Howes:2011a} but sufficiently low that the measured
damping rate of the mode agrees with the collisionless value.  For the
linear relaxation phase, we apply an enhanced collisionality with a
value $\nu_s = 10 \gamma_s$. The linear relaxation continues until a
simple exponential damping rate $\gamma$ for the mode is achieved.
After completion of the linear relaxation, tests show that the
resulting mode has a frequency and damping rate in close agreement
with the prediction of the linear, collisionless dispersion relation.
These modes are then used to begin the nonlinear simulation of the
counterpropagating \Alfven wave collision. Note that, for \Alfven
waves in the MHD limit, for which collisionless damping is weak,
$\gamma \ll \omega_r$, even the enhanced collisionality of the linear
relaxation phase corresponds to weakly collisional dynamics since $\nu
\ll \omega_r$.

%

\end{document}